\documentstyle[prl,aps,epsf]{revtex}
\draft
\begin{document}
\twocolumn[\hsize\textwidth\columnwidth\hsize\csname @twocolumnfalse\endcsname

\title{Disordering Transitions in Vortex Matter: Peak Effect and
Phase Diagram}
\author{C. J. Olson, C. Reichhardt, 
R. T. Scalettar, and G. T. Zim{\' a}nyi}
\address{Department of Physics, University of California, Davis, California
95616.}
\author{Niels Gr{\o}nbech-Jensen} 
\address{Department of Applied Science, University of California, Davis,
California 95616.\\
NERSC, Lawrence Berkeley National Laboratory, Berkeley, California 94720.} 

\date{\today}
\maketitle
\begin{abstract}
Using numerical simulations of magnetically interacting
vortices in disordered layered superconductors we obtain the static
vortex phase diagram as a function of magnetic field and temperature.
For increasing field or temperature, we find a
transition from ordered straight vortices to disordered decoupled 
vortices. This transition is associated with 
a peak effect in the critical current. 
For samples with increasing disorder strength
the field at which the decoupling occurs decreases.
Long range, nonlinear 
interactions in the c-axis are required to observe the effect.
\end{abstract}
\vspace{-0.1in}
\pacs{PACS numbers: 74.60.Ge, 74.60.Jg}
\vspace{-0.3in}
\vskip2pc]
\narrowtext

Vortex matter in superconductors exhibits
a remarkably rich variety of distinct phases 
due to the numerous competing interactions \cite{1}.  
These phases can strongly affect the response of the
system, such as the critical current $J_c$ and magnetization. 
In highly anisotropic superconductors, such as 
BSCCO, a pronounced fish-tail or peak effect,
in which $J_c$ shows a sharp increase, is
observed as a function of increasing field \cite{2}. 
This peak can be interpreted as occurring when the
increasing field weakens the interlayer coupling of vortex pancakes
due to geometric constraints,
and a transition occurs from weakly pinned 3D line vortices to 
decoupled 2D pancake vortices which can more easily adjust their positions
to maximize the pinning \cite{3}. 
The peak has also been proposed to arise from plasticity, proliferation
of in-plane defects, dynamical effects, or matching effects
\cite{4,5,6}.

There is mounting experimental evidence that the peak effect is
associated with a sharp transition in the vortex lattice from an
ordered state to a disordered state. In BSCCO, 
neutron scattering \cite{7} and muon lifetime \cite{8} experiments
provide evidence that a 
transition from an ordered 3D vortex arrangement to a  
disordered or decoupled arrangement is associated with the peak effect.
Additional evidence from plasma-resonance \cite{9} 
and magneto-optical studies \cite{10}
point to the first-order nature of this transition.
In YBCO a rapid increase in $J_c$ as a function of magnetic 
field is observed and is thought to indicate a transition from an ordered
to a disordered state. History and memory effects
found near this peak indicate 
that this transition is also first order \cite{5},
suggesting that the physics of the second peak is similar in BSCCO and YBCO.  
In recent 
muon-spin measurements in YBCO, the order-disorder transition has been
interpreted
to be a 3D-2D transition of the vortex lattice \cite{11}. 
As a function of temperature, in YBCO a peak effect can be observed very near
or at the melting line \cite{12,13}. 
Transformer measurements in this regime
provide evidence of  vortex cutting in the liquid state,
suggesting that the breakdown of the 3D nature of
vortex lines is relevant in this case as well.     

A key question is what type of sharp transition is responsible for
the observed peak in $J_c$, and whether
the mechanism of the peak effect is the same going through the 
melting line as through the second peak.        
In this paper we argue that it is a decoupling transition along the
vortex line, in the $c$-direction, which is responsible.  We present
results from a simulation of magnetically interacting vortices
in which we demonstrate a robust peak in $J_c$ 
as a function of magnetic field and produce an
H-T phase diagram in good agreement with experiment.

We consider a 3D layered superconducting model material containing an
equal number of pancake vortices in each layer, interacting magnetically.
This simulation differs from those used previously by several other groups
in that it treats the long-range interactions along the length of the
vortex line exactly, and neglects the Josephson coupling, approximating 
highly anisotropic materials.
Previous models have treated the inter-plane interactions as
simple (linear) elastic connections 
\cite{14,15,16,17}.

The overdamped equation of motion for vortex $i$ is 
$ {\bf f}_{i} = -\sum_{j=1}^{N_{v}}\nabla_{i} {\bf U}(\rho_{ij},z_{ij})
+ {\bf f}_{i}^{vp} + {\bf f}_{d} + {\bf f}^{T}= {\bf v}_{i}$. 
The total number of 
pancakes per layer
is $N_{v}$, and $\rho_{ij}$ and $z_{ij}$ are the distance between
vortex $i$ and vortex $j$ in cylindrical coordinates. We impose 
periodic boundary conditions in $x$ and $y$ directions \cite{18}.
The interaction energy between pancakes is \cite{19,20}
\begin{eqnarray}
{\bf U}(\rho_{ij},0)=2d\epsilon_{0}
\left((1-\frac{d}{2\lambda})\ln{\frac{R}{\rho}}
+\frac{d}{2\lambda}
E_{1}\right) \ ,
\nonumber
\end{eqnarray}
\begin{eqnarray}
{\bf U}(\rho_{ij},z)=-s_{m}\frac{d^{2}\epsilon_{0}}{\lambda}
\left(\exp(-z/\lambda)\ln\frac{R}{\rho}+
E_{2}\right) \ , \nonumber
\end{eqnarray}
where $R$ is
the maximum in-plane distance, 
$E_{1} = 
\int^{\infty}_{\rho} d\rho^{\prime} \exp(\rho^{\prime}/\lambda)/\rho^{\prime}$,
$E_{2} = 
\int^{\infty}_{\rho} d\rho^{\prime} \exp(\sqrt{z^{2}+\rho^{\prime 2}}/\lambda)/\rho^{\prime}$,
$\epsilon_{0} = \Phi_{0}^{2}/(4\pi\lambda)^{2}$, $d=0.005\lambda$ is the
interlayer spacing,
and $\lambda$ is the London penetration depth. 
We model the pinning
as $N_p$ short range attractive 
parabolic traps that are randomly distributed in each 
layer.
The pinning interaction is 
${\bf f}_{i}^{vp} = -\sum_{k=1}^{N_{p}}(f_{p}/\xi_{p})
({\bf r}_{i} - {\bf r}_{k}^{(p)})\Theta( 
(\xi_{p} - |{\bf r}_{i} - {\bf r}_{k}^{(p)} |)/\lambda)$,
where the pin radius is $\xi_{p}$,
the pinning force is $f_{p}$, and $f_{0}^{*}=\epsilon_{0}/\lambda$.
Thermal fluctuations are represented as Gaussian noise of width
$f_T$, with
$<f^{T}_{i}(t)>=0$ and 
$<f^{T}_{i}(t) f^{T}_{j}(t^{\prime})> = 
2\eta k_B T \delta_{ij} \delta(t-t^{\prime})$.
The parameter $s_m$ is used to vary the coupling strength between
planes, but except where noted it is set to $s_m=1.0$.
To vary the applied magnetic field $H$, we fix the number of
vortices in the system and change the system size (vortex density $n_v$).
The pin density remains fixed, but in
all cases $N_p>N_v$.
We consider systems of $L=8$ to 32 layers containing from 
1 to 80 vortex pancakes
per layer.  The system size ranges from $0.6\lambda \times 
0.6\lambda$ to $141\lambda \times 141\lambda$.

The pancakes in our model can behave in two possible ways.
They may align along the c-axis into well-defined vortex lines (``3D'').
Alternatively, the pancakes may break apart in the c-direction and 
move independently in each plane (``2D'').
We can cross between the two types of behavior in the presence of
disorder by varying the inter-plane coupling strength, $s_m$.
The vortices are 3D at high coupling, while 2D at low coupling.  
We find that due to the long range interactions and the nonlinearity, the
transition between these two states is sharp and has many first-order
characteristics, including strong hysteresis as well as superheating and
supercooling effects.  We have studied this coupling strength crossover
in detail in previous papers \cite{21}, 
and have shown that it is associated with
a large change in the critical current $f_c$, as indicated in Fig.\ 1(a).
To quantify the alignment of the vortices between layers, we also
plot 
the correlation function in the $z$-direction,  
$C_{z} = 1 - \langle\Theta(a_{0}/2-|{\bf r}_{i,L}-{\bf r}_{j,L+1}|)
|{\bf r}_{i,L}-{\bf r}_{j,L+1}|2/a_{0}\rangle$, where $a_{0}$ is the
vortex lattice constant. When $C_z=1.0$, the pancakes are aligned into
3D lines, whereas a low value of $C_z$ indicates that the pancakes
are decoupled.

It is expected that changing the magnetic field in the system will cause
a decoupling transition for the same reason that changing the coupling
strength for interlayer interactions does.
This is because the vortices come closer together in plane 
as the field increases while the spacing between planes
remains constant.  Thus, at higher fields the relative coupling
between planes becomes weaker. 
Here, we 
show that the same decoupling transition previously 
observed as a function of coupling strength also occurs 
as a function of magnetic field. 

In Fig.\ 1(b-d) we plot $f_c$
as a function of
vortex density (magnetic field) $n_v$,
at a fixed temperature of $f_{T}=0.0005$
for several different samples containing different
numbers of vortices.  A clear peak or 
fish-tail effect is present.  
In Fig.\ 1(c), a sample containing $N_v=4$ is shown.  Here,
for low fields ($n_v<0.1$), 
$f_c$ is high and the vortices are
uncorrelated in the $z$ direction as well as disordered in plane. 
At these low fields the vortex-pin interactions dominate.
$f_c$ drops rapidly with field in this regime
due to the 

\begin{figure}
\center{
\epsfxsize=3.5in
\epsfbox{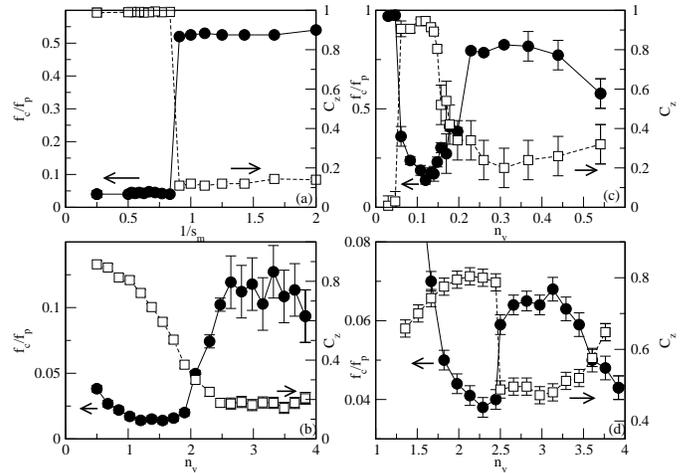}}
\caption{(a) 3D-2D transition 
as a function of 
inverse interlayer coupling strength, $1/s_m$, for a sample in
which $N_v=80$, $L=16$, $n_v=0.35$, $n_p=1.0$, and $f_p=0.02f_{0}^{*}$.
Filled circles: $f_c$; open squares: $C_z$.
(b-d) $f_c$ (filled circles) and $C_z$ (open squares) as a function
of vortex density $n_v$ in samples with $s_m=1.0$ and:
(b) $N_v=1$, $L=32$, $n_p=40.0$, and $f_p=0.2f_{0}^{*}$.
(c) $N_v=4$, $L=8$, $n_p=2.0$, and $f_p=0.04f_{0}^{*}$;
(d) $N_v=80$, $L=16$, $n_p=8.0$, and $f_p=0.1f_{0}^{*}$.
}
\label{fig:fig1}
\end{figure}

\hspace{-13pt}
increasing in-plane vortex lattice stiffness, which causes
the pinning to be less effective \cite{22}.
For intermediate fields, $0.1<n_v<0.2$, $f_{c}$ is low  
and the vortices form an ordered 3D structure as indicated by a near unity
correlation, $C_{z} \sim 1$. 
The 3D vortex lines in this regime are poorly pinned by the point pinning.
At $n_v = 0.1$ there is a sharp 
decoupling transition as is apparent in the 
abrupt drop of $C_{z}$. 
Simultaneously, $f_{c}$ increases rapidly to $f_c = 0.8f_p$. 
At higher fields $f_{c}$ gradually decreases 
due to the increasing in-plane interactions, which continue to
stiffen the vortex lattice and thereby weaken the effectiveness of the pinning.
We have observed similar peaks in the critical current for samples
containing $N_v=12$, $30$, and $80$ vortices.  Data for the $N_v=80$ 
sample is shown in Fig. 1(d).  The exact value of $n_v$ at which the
peak falls is affected by the pinning strength and can be adjusted
within a wide range, as indicated in Fig. 1.

To show that the peak effect observed here
arises from the onset of plasticity
of the vortices {\it between} layers, and not from {\it in-plane}
plasticity,
we conduct simulations of a rigid vortex lattice, represented by
a single vortex with periodic boundary conditions.
In the $N_v=1$ simulation in-plane defects are not permitted.
Thus, the vortex line can disorder along the c-axis only.
The fact that we can observe the second peak in a sample 
containing
only a rigid lattice, as in Fig.\ 1(b), indicates that the observed
peak is not being generated by the proliferation of defects in the plane.
This, along with our previous results on the existence
of a transition as a function of $c$-axis coupling strength,
provides evidence that it is a change in $c$-axis correlation that causes
striking changes in $f_c$.

We can now probe the effect of temperature on the 

\begin{figure}
\center{
\epsfxsize=3.5in
\epsfbox{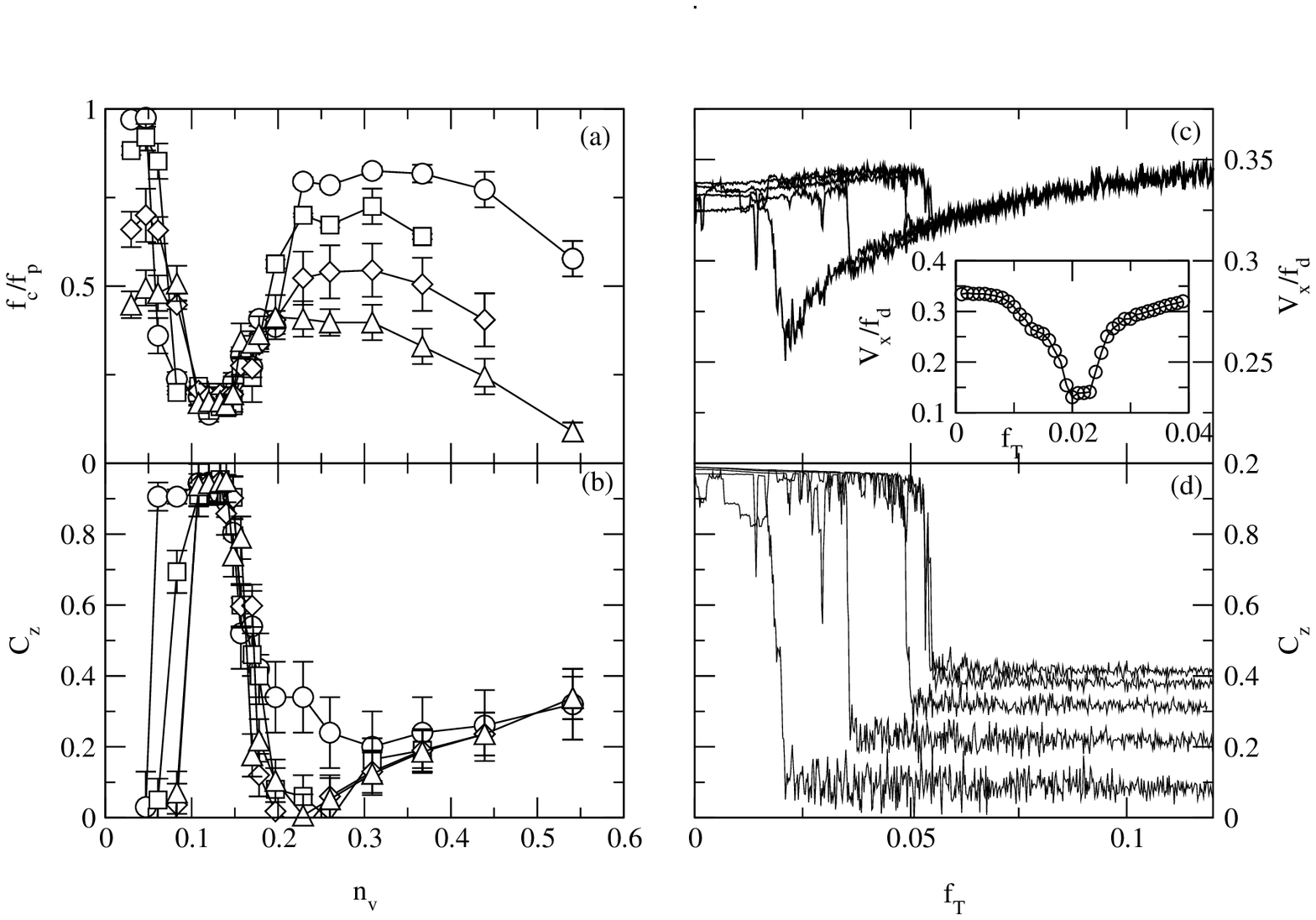}}
\caption{(a) $f_c$ versus $n_v$ for a sample with $N_v=4$, $L=8$, and
$f_p=0.04f_{0}^{*}$ 
at temperatures of $f_T=0.0005$ (circles), $f_T=0.1$ (squares),
$f_T=0.2$ (diamonds), and $f_T=0.3$ (triangles).  
(b) $C_z$ corresponding to panel (a).
(c) $V_x$ versus $f_T$ for the same sample, at a fixed drive
of $f_d=0.1f_{0}^{*}$, for fields $n_v=$ 0.061, 0.083, 0.108, 0.133, and 0.178.
(d) $C_z$ corresponding to panel (c). Inset to (c): $V_x$ versus $f_T$
for a sample with $N_v=80$, $n_{v}=2.0$, 
$L=16$, $n_{p}=8.0$, and $f_{p}=0.1f_{0}^{*}$.}
\label{fig:fig2}
\end{figure}

\hspace{-13pt}
observed transition.
As illustrated in Fig.\ 2(a-b), the field
at which the 3D-2D transition occurs remains almost constant 
with $T$, in agreement with experiment \cite{23}.
$f_c$ in the decoupled state drops as the
temperature is raised,
but the jump in $f_c$ persists. 
In Fig.\ 2(c) we show the peak effect in 
the average vortex velocity in the direction of drive, $V_x=<v_x>$, 
as a function of
temperature for fixed values of $n_v$.  Starting at $f_T = 0$, we
apply a constant driving force of $f_{d} = 0.1f_{0}^{*}$ such that
the vortices are moving and ordered.
As $f_T$ 
is increased there is a transition to a 2D decoupled vortex arrangement
as seen in the drop in $C_{z}$.  This 
coincides with a sharp drop in the vortex velocities $V_x$
since the decoupled vortices experience stronger effective 
pinning. As $f_T$ is further increased, the effectiveness of
the pinning gradually decreases and $V_x$
gradually increases back to the free flow level.   
The $f_c$ measurements at different temperatures 
and the resistivity measurements
as a function of temperature 
exhibit the same features as
experimental data
taken near the peak effect regime.

Next, we construct a phase diagram for our model in the H-T plane.
Since we wish to stress that all qualitative features of the phase diagram
result from the c-axis interactions, we consider
a system containing a rigid vortex lattice.
We first locate the thermal melting line for $N_v=1$ in
the absence of pinning.  This is shown as open circles on the
phase diagram.  We find a reentrance in 
the melting, shown
in more detail in the top inset of Fig.\ 3, which agrees
with theoretical predictions for the melting \cite{24}.
In the presence of pinning, we measure $f_c$
as a function of field (as in Fig.\ 1) at several different temperatures.
We plot the 3D-2D transition field as solid squares.
As was shown in Fig.\ 2, the location of the transition is
insensitive to temperature at low 

\begin{figure}
\center{
\epsfxsize=3.5in
\epsfbox{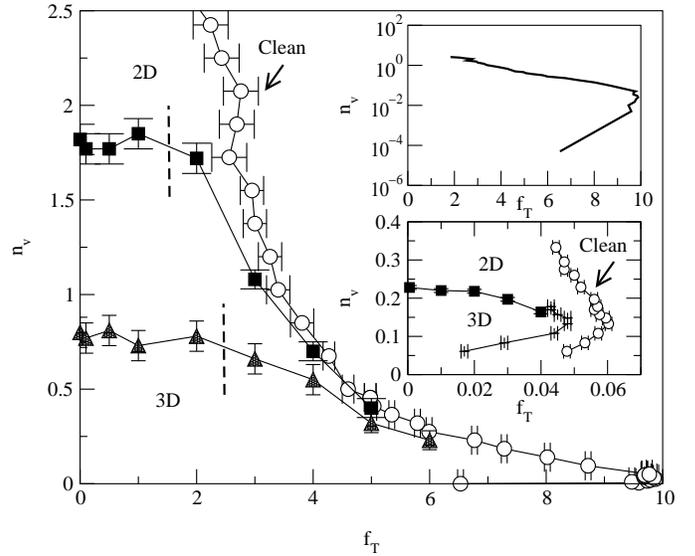}}
\caption{Phase diagram for a sample with $N_v=1$ and $L=32$.
Open circles: decoupling transition for $f_p=0.0$
(clean melting line).  Filled squares: decoupling
transition for a sample with $f_p=0.2f_{0}^{*}$.  Triangles: decoupling
for a sample with $f_p=1.6f_{0}^{*}$.
Upper inset: Reentrance in the clean melting line is shown on a log-log
plot.
Lower inset: Phase diagram for a sample with $N_v=4$ and $L=8$, with
stronger pinning of $f_p=0.04f_{0}^{*}$.  
Circles: clean melting line.  Squares: decoupling transition
determined from $f_c$ vs. $n_v$ data as in Fig.\ 2(a).  Plus signs:
decoupling transition determined from $V_x$ vs. $f_T$ data as in 
Fig.\ 2(c).}
\label{fig:fig3}
\end{figure}

\hspace{-13pt}
temperature.  We find,
however, that the temperature weakens the pinning enough that for
$f_T>1$ a transition in $f_c$ is no longer observed.
Instead $f_c$ is small everywhere.  The transition is still
observable in $C_z$, 
and data taken from $C_z$ is shown above
$f_T=1$, indicating that the transition begins to drop in field at
higher temperatures until it merges with the clean melting line.

The field at which the second peak occurs is lowered in samples with
stronger pinning,  as indicated in Fig.\ 3 where the second peak
line for a sample with $N_v=1$ and stronger pinning is plotted.
This is consistent with experiments in which the peak
shifts to lower fields in samples 
that have had their pinning strength increased artificially.

We have also constructed a phase diagram for a sample with $N_v=4$
and much stronger  pinning, shown in the lower inset of Fig.\ 3.  
In the disordered system the  
3D-2D transition is relatively flat in $f_T$ for $f_T< 0.3$ and then
begins to turn down as the clean melting line is approached. We also 
observe that the clean system shows a reentrant behavior for low 
fields due to the reduced vortex interactions. 
In the disordered sample this reentrant line is pushed up in field.
We see reentrance of
the second peak line in good agreement with recent measurements
\cite{23}.
In still larger systems containing $N_v=80$ vortices, we have taken
two slices across the phase diagram as a function of field and 
temperature,
shown in Fig. 1(d) and 

\begin{figure}
\center{
\epsfxsize=3.5in
\epsfbox{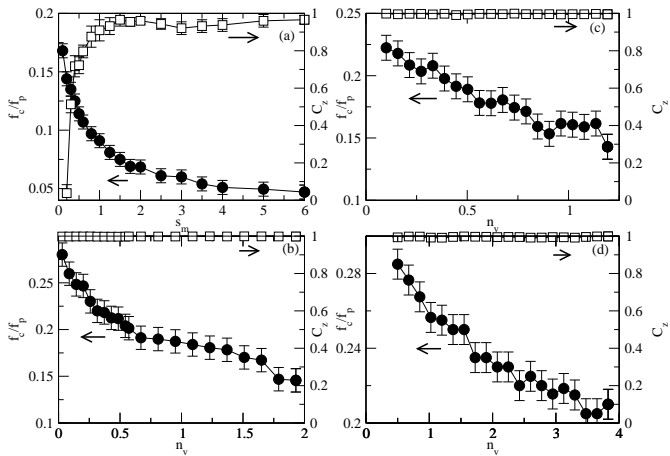}}
\caption{Critical current as a function of magnetic field for a model
containing springs between layers.  
(a) $N_v=1$, $L=32$, $f_p=0.2f_{0}^{*}$. (b) $N_v=4$, $L=8$, 
$f_p=0.04f_{0}^{*}$.
(c) $N_v=12$, $L=16$, $f_p=0.1f_{0}^{*}$. (d) $N_v=80$,
$L=16$, $f_p=0.1f_{0}^{*}$.}
\label{fig:fig4}
\end{figure}

\hspace{-13pt}
the inset of Fig. 2(c), respectively, and find decoupling transitions consistent with those
shown above.

To show that it is the long range interactions and the
nonlinearity in the c-axis that
cause the sharpness of the transition, we 
replace the long range interplane interactions 
by a model in
which the pancakes are coupled by elastic nearest-neighbor springs.  This
model resembles one considered earlier \cite{16,17}.
The interlayer interaction is 
${\bf f}_{i,j} = -s_{m}({\bf r}_{i,L+1} - 2{\bf r}_{i,L}+{\bf r}_{i,L-1})$,
where $s_m$ can be used to vary the coupling strength.

As shown in Fig.\ 4, for varying numbers of vortices, we find a
{\it smooth} change in depinning force 
for the same parameters 
and over the entire range of $n_v$ 
considered in the long range interacting case.
For $N_v=1$, the interlayer interaction of the spring is not
affected by varying $n_v$ so we vary $s_m$ instead.  
For $N_v=4$, 12, and 80, we set $s_m=1$ and vary $n_v$ directly.
In each case we find a smooth
decrease of $f_c$ with field.  
We do not observe any indications that a model containing springs and
lacking long range interaction and nonlinearity in the z-direction can show
a statistically significant peak in the critical current, as was recently 
suggested \cite{17,25}.

In conclusion,
using numerical simulations of magnetically interacting vortex pancakes in
three dimensional layered superconductors, we have obtained static vortex
phase diagrams as a function of magnetic field and temperature.
We have demonstrated a unique
relationship between c-axis vortex correlations and the critical current
$J_c$.
Specifically, we have identified c-axis correlation transitions to be
responsible for the peaks in $J_c$ as well as the fish-tail
effect in magnetization measurements. 
Our simulations of rigid in-plane vortex
lattices, which also exhibit the peak and fish-tail phenomena,
support the claim that changes in
c-axis (as opposed to in-plane) correlations are responsible for the
observed anomalies. 
Finally, we have addressed recent suggestions that the 
peak effect in $J_c$ 
can be observed in a model system where inter-plane interactions
are modeled by linear springs between pancakes. We 
only observe the critical current peak when long-range 
interactions are present,
again suggesting that the
peak and fish-tail effect are directly related to the nonlinear and nonlocal
nature of the inter-plane vortex interactions that lead to decoupling
transitions.

We acknowledge useful discussions with D.\ Dom\'{\i}nguez, A.\ Kolton,
A. Koshelev, W.K. Kwok, V. Vinokur, and E. Zeldov.
This work was supported by CLC and CULAR (LANL/UC), by NSF-DMR-9985978,
and by the Director, Office of Adv.\ Scientific Comp.\ Res., Div.\ of Math.,
Information and Comp.\ Sciences, U.S.\ DoE contract
DE-AC03-76SF00098.


\begin{references}
\bibitem{25} Indeed, we have recently found that
 the peak reported in Ref.\ \cite{17} is not robust to
 new and more extensive simulations which use 
 a larger number of disorder realizations.
\end{references}
\end{document}